\documentstyle [12pt] {article}
\begin{document}
\pagestyle{empty}
\textwidth 6truein
\textheight 8.5truein
 \pagenumbering{arabic}

\centerline {\bf Symmetric Solutions of}
\centerline {{\bf Einstein-Yang-Mills Equations}
 \footnote{to appear in the Proceedings of the
 \it First Iberian Meeting on
 Gravity, \rm \'{E}vora, Portugal, September, 1992;
 World Scientific, 1993}
}

\vskip 0.5cm

\centerline{ JOS\'E MOUR\~AO \footnote{e-mail: JMOURAO@PTIFM.bitnet}}

\centerline{\it Departamento de F\'{\i}sica, Instituto Superior T\'{e}cnico,
Lisboa, Portugal}

\vskip 1cm

\centerline{\bf Abstract}

Symmetric gauge fields and invariant metrics in homogeneous
spaces are found. Their use for finding
exact solutions of the Einstein-Yang-Mills (EYM) equations
is discussed.

\bigskip

\vskip 1cm

\noindent {\bf 1. Introduction}

\vskip 0.5cm

The search for exact solutions in general relativity
can usually be divided in two parts. The first part, which we
call "kinematical", consists in the restriction to a subset
of fields in the space of field configurations (i.e. in the
making of appropriate ans\"{a}tze).

This  restriction is made in accordance with the physical problem
in consideration and is achieved by considering only the
configurations which are invariant (or symmetric) under the
action of a symmetry group $S$.
For example in the case of
 closed, flat and open Friedmann-Robertson-Walker
cosmological models, $S$ is respectively
$SO(4)$,
 $E^3$,
$ SO(1,3)$
and plays the role of group of spatial
homogeneity and isotropy.

In the second (or "dynamical") part of the search for exact solutions
in General Relativity we study the equations
implied by the original theory
in the set of $S$-symmetric configurations.
The number of independent variables
in these equations is equal to the dimensionality of space-time minus
the dimensionality of the orbits of the symmetry group.

Here we will describe,
following [1,2], a method of finding invariant metrics and
symmetric gauge fields
used in the
 kinematical part of the search for solutions of
the EYM equations  (see also [3,4]).

\vskip 0.5cm

\noindent
{\bf 2. Invariant Metrics and  Symmetric Gauge Fields}

\vskip 0.5cm

Consider a space-time of the form
$$
M = \tilde M \times S/R   \ \ ,    \eqno(1)
$$
where $S$ and $R$ are Lie groups,
and assume that we are interested in $S$-symmetric solutions of the
EYM equations.
The gauge group $K$ is assumed to be a compact Lie group.

Let us for further simplicity restrict ourselves to
a homogeneous space-time
$$
M = S/R \ \ .   \eqno(2)
$$
The reader interested in the general case (1) is referred to the
contribution by Kapetanakis and Zoupanos in the present volume and to the
literature [1-5]. Notice that the $S$-invariant ansatz
for the metric and the $S$-symmetric (i.e. invariant up to a gauge
transformation) ansatz for the gauge field, discussed here, are
always present in the general case.

We first chose a prefered moving frame in $S/R$ and
describe the $S$-invariant ansatz for the metric
in $S/R$.

Consider the canonical (${\cal S} \equiv  Lie(S)$-valued) left-invariant
one-form on the group manifold of S
$$
\theta = \theta^{\alpha} T_{\alpha} = s^{-1} d s    \    ,
$$
where the $T_{\alpha}$ are the generators of $S$ and
 $ \theta^{\alpha}$ are left-invariant one-forms on $S$.
The form $\theta$ is very important for the study of the
invariant geometry of the
Lie group $S$ and its pull back to $S/R$
plays, as we shall see below,
 an analogous role in the study of the invariant geometry
of the coset space $S/R$.
The group $S$ can be considered as a principal bundle over
$S/R$ with structure group $R$ (we assume $R$ to be a closed subgroup
of $S$
 and  $S/R$ to be a reductive coset space).
Let us chose a (local) section $\sigma$ of this bundle
 in a neighborood of the origin
$o \equiv [e ] \equiv R$
$$
\sigma : {\cal U} \subset  S/R  \longrightarrow S
$$
and pull the form $\theta$ back to $S/R$ with respect to $\sigma$:
$$
\bar \theta_x = \left( \sigma^* \theta \right)_x = \sigma^{-1}(x) d \sigma (x)
= \bar \theta^\alpha T_\alpha \ \ . \eqno(3)
$$
The $\cal S$-valued one-form $\bar \theta$ in (3)
is called the Maurer-Cartan one-form in $S/R$. A local moving
coframe  can be obtained from the Maurer Cartan one form as
follows. Let
$$
{\cal S} = {\cal R} + {\cal M}      \eqno(4)
$$
be a reductive decomposition of $\cal S$, where ${\cal R} \equiv Lie(R)$
and $\cal M$ is a reductive
(i.e.
$[ {\cal R} , {\cal M}] \subset  {\cal M}$) complementary subspace
and
$$
\{ T_\alpha \} = \{T_a, T_{\dot a} \} \ \ , T_{ a} \in {\cal M}, \
T_{\dot a} \in {\cal R}     \eqno(5)
$$
be an adapted basis in $\cal S$. We have the natural decomposition
of the Maurer-Cartan one-form associated with (5)
$$
\bar \theta = \bar \theta_{\cal M} + \bar \theta_{\cal R}
\equiv \bar \theta^a T_a +  \bar \theta^{\dot a} T_{\dot a} \eqno(6)
$$
Unlike the group case $S$,
the one-forms $\bar \theta^\alpha$  in $S/R$ are, in general, not
$S$-invariant. Nevertheless they have simple transformation laws
under the action of $S$. For given $s \in S, x \in {\cal U}$ and $\sigma$,
let $r_s(x; \sigma)$ be the element
of the isotropy group $R$
defined by
$$
\sigma (s x) = s \sigma (x) r_s (x ; \sigma)   \ \  .
$$
Then the transformation law of
$\bar \theta^a$ and $\bar \theta^{\dot a} $
under the action of $s \in S$
can be easily obtained
from
$$
s^* \bar \theta_{\cal M} = ad \ r_s^{-1}(x) \bar \theta_{\cal M}  \eqno(7a)
$$
$$
s^* \bar \theta_{\cal R} = ad \ r_s^{-1}(x) \bar \theta_{\cal R}
+ r_s^{-1}(x) d r_s(x)     \    \    ,
 \eqno(7b)
$$
where $ad$ denotes the adjoint representation of $S$.
The reductiveness of the decomposition (4) implies that the restriction
of $ad$ to the subgroup $R$ has
 $\cal M$  as an invariant subspace. The representation of $R$ in
 $\cal M$
obtained in this way is called the isotropy
representation and is denoted by $ad_R$. Let $\{ X_a \}$ be the frame dual
to $\{ \bar \theta^a \}$. It is clear from (7a) that
under $s_*$ the vectors $X_a$ in this frame transform in the same way
 the $T_a$
do under
the isotropy action of $ r_s^{-1}(x)$ (i.e. $ad \ r_s^{-1}(x)$).
Let $B(.,.)$, be an $ad_R$-invariant scalar product in $\cal M$ with
components $B_{ab} = B(T_a, T_b)$. Then we conclude
[6] that the metric
in $S/R$ with the same components $B_{ab}$
in the frame $\{ X_a \}$,
$$
\gamma = B_{ab} \bar \theta^a \otimes \bar \theta^b       \eqno(8)
$$
is $S$-invariant. Moreover any $S$-invariant metric in $S/R$ can be
constructed in this way.
Therefore we have succeded in reducing the
(differential geometry) problem of finding $S$-invariant
metrics in the coset space $S/R$
to the purely algebraic problem of finding $ad_R$-invariant scalar
products in $\cal M$. The equation (8) with $B_{ab}$ being
an $ad_R$-invariant scalar product in $\cal M$ gives the ansatz
for $S$-invariant metrics in $S/R$.

In an analogous way [1-5] the problem of finding $S$-symmetric gauge
fields with gauge group $K$ in $S/R$ is reduced by Wang's theorem [6] to
the algebraic problem of intertwining equivalent representations
of the isotropy group. To see this recall that, according to
Wang's theorem, the most general form of a $S$-symmetric
gauge field in $S/R$ is given by
$$
A = \lambda(\bar \theta_{\cal R}) + \Lambda (\bar \theta_{\cal M}) \ \ ,
\eqno(9)
$$
where
$$
\lambda: {\cal R} \longrightarrow {\cal K} = Lie (K)     \eqno(10a)
$$
is an homomorphism from the isotropy algebra $\cal R$ to the
Lie algebra of the gauge group $\cal K $
and $\Lambda$ is a mapping from $\cal M$ to $\cal K$
$$
\Lambda : {\cal M} \longrightarrow {\cal K}   \eqno(10b)
$$
satisfying the linear property
$$
 \Lambda(ad(r) u) = ad (\lambda(r)) \Lambda(u)   \ \ ,
\qquad u \in {\cal M} \ \ , \qquad r \in R  \ \
 \eqno(10c)
$$
This property  means that the mapp $\Lambda$ intertwines the
representation of $R$ in $\cal M$ (i.e. the isotropy
representation) with the representation of $\lambda(R)$ in
$\cal K$ (obtained by restricting the adjoint representation
of $K$ to $\lambda(R)$). The solutions of (10c) are therefore given
by Schur's lemma.

By making  the ans\"atze (8) and (9, 10)
we have concluded the kinematical
part of the search for $S$-symmetric solutions of the EYM equations.

Analagous symmetric ans\"atze for linear connections with torsion
have been made in the framework  of generalized
Einstein-Cartan theories [7].

\vskip 0.5cm

\noindent
{\bf 3. Acknowledgements}

\vskip 0.5cm

It is a pleasure to thank M.C. Bento, O. Bertolami, A.B. Henriques,
Yu. A. Kubyshin, P.V. Moniz, R. Picken, P. S\'{a}, J.N. Tavares and
I.P. Volobuev for collaboration over the years and Yu. A. Kubyshin for
usefull suggestions.

\vskip 0.3cm
\noindent
\bf 4. References \rm
\noindent
\begin{itemize}
\item[1.] I.P. Volobuev and Yu.A. Kubyshin,
 \it JETP Lett.
\rm \bf 45 \rm (1987) 581; \it Theor. Math. Phys.
\rm \bf 75 \rm (1988) 509;

Yu.A. Kubyshin, J.M. Mour\~{a}o, G. Rudolph and I.P. Volobuev,
 \it Lect. Notes in Physics
\rm \bf 349 \rm (Springer Verlag, 1988);

Yu.A. Kubyshin, J.M. Mour\~{a}o and I.P. Volobuev,
 \it Phys. Lett.
\rm \bf B203 \rm (1988) 349;
\it Nucl. Phys.
\rm \bf B322 \rm (1989) 531

\item[2.]  O. Bertolami, J.M. Mour\~{a}o, R. Picken and I.P. Volobuev,
 \it Int. J. Mod. Phys.
\rm \bf A6 \rm (1991) 4149;

P.V. Moniz and J.M. Mour\~{a}o,
 \it Class. Quant. Grav.
\rm \bf 8 \rm (1991) 1815;

O. Bertolami and J.M. Mour\~{a}o,
\it Class. Quant. Grav.
\rm \bf 8 \rm (1991) 1271;

O. Bertolami, Yu.A. Kubyshin and J.M. Mour\~{a}o,
 \it Phys. Rev. D
\rm \bf 45 \rm (1992) 3405;

P.V. Moniz, J.M. Mour\~{a}o and P.M. S\'{a},
 \it Class. Quant. Grav.
 \rm in Press;

M.C. Bento, O. Bertolami, P.V. Moniz, J.M. Mour\~{a}o and P.M. S\'{a},
 \it Class. Quant. Grav.
 \rm in Press

\item[3.] M. Henneaux,
 \it J. Math. Phys.
\rm \bf 23 \rm (1982) 830;

Yu.A. Kubyshin, V.A. Rubakov and I.I. Tkachev,
 \it Int. J. Mod. Phys.
\rm \bf A4 \rm (1989) 1409;

R. Coquereaux and A. Jadczyk,
 \it Lect. Notes in Physics
\rm \bf 16 \rm (World Scientific, 1988);

  K. Yoshida, S. Hirenzaki and K. Shiraishi,
 \it Phys. Rev. D
\rm \bf 42 \rm (1990) 1973

Yu. A. Kubyshin, J.I. Perez Cadenas,
\it  Preprint Nuclear Phys. Inst., Moscow State University, \rm
Moscow (1993)

\item[4.] H.P. Kunzle,
 \it Class. Quant. Grav.
\rm \bf 8 \rm (1991) 2283

\item[5.] P. Forgacs and N.S. Manton,
 \it Comm. Math. Phys.
\rm \bf 72 \rm (1980) 15

I.P. Volobuev and Yu.A. Kubyshin,
 \it  Theor. Math. Phys.
\rm \bf 68 \rm (1986) 788;
\rm \bf 68 \rm (1986) 885

F.A. Bais, K.J. Barnes, P. Forgacs and G. Zoupanos
 \it  Nucl. Phys.
\rm \bf B291 \rm (1986) 557;

G. Rudolph and I.P. Volobuev,
 \it  Nucl. Phys.
\rm \bf B313 \rm (1989) 95;

D. Kapetanakis and G. Zoupanos,
 \it  Phys. Lett.
\rm \bf B249 \rm (1990) 73;
 \it  Phys. Rep.
\rm \bf C 219 \rm (1992) 1;

S. Shabanov,
 \it  Phys. Lett.
\rm \bf B272 \rm (1992) 11

\item[6.] S. Kobayashi and K. Nomizu, "Foundations of
Differential Geometry"  (Interscience Publishers, 1969)

\item[7.] Yu. A. Kubyshin, O. Richter and G. Rudolph,
 \it  Rep. Math. Phys.
\rm \bf 30 \rm (1991) 355;
\it Leipzig
Univ. Preprint, \rm Leipzig (1992)

Yu. A. Kubyshin, V. Malyshenko and D. Marin Ricoy,
 \it  Preprint Nuclear Phys. Inst., Moscow State University, \rm
Moscow (1993)

\end{itemize}
\end{document}